\documentclass[letterpaper,12pt]{article}   

\usepackage{amsfonts}
\usepackage{graphicx}
\usepackage{epsfig}
\usepackage{graphics}

\makeatletter
\newcommand{\row}[1]%
{\mathord{\buildrel{\lower3pt%
\hbox{$\scriptscriptstyle\rightarrow$}}\over #1}}

\newcommand{\dyadic}[1]{\mathord{\dyadic@rrow{#1}}}
\newcommand{\dyadic@rrow}[1]{
\begin{picture}(12,12)(-1,0)
\put(-3,12){\makebox(0,0)[t]{$\scriptscriptstyle\downarrow$}}
\put(-3,13){\makebox(0,0)[l]{$\scriptscriptstyle\longrightarrow$}}
\put(5,0){\makebox(0,0)[b]{$#1$}}
\end{picture}
}

\newcommand{\bra}[1]{\bigl\langle #1 \bigr|}
\newcommand{\ket}[1]{\bigl| #1 \bigr\rangle}

\input{tcilatex}

\begin{document}
\begin{center}
{\large Some Properties of  Multi-Qubit States  in Non-Inertial
Frames}

\vspace{0.5cm}
\renewcommand{\thefootnote}{\fnsymbol{footnote}}
 Alaa Sagheer\footnote{Corresponding author}$^{\dagger \ddagger} $ and Hala Hamdoun$^\ddagger$\\
$^\dagger$Department  of Mathematics\\
$^\ddagger$Center for Artificial Intelligence and RObotics (CAIRO)\\
Faculty of Science, Aswan University, Aswan,
Egypt\\
Email: \{alaa, hala\}@cairo-svu.edu.eg\\

\end{center}
{\bf Abstract}: In this paper, some properties of multi-qubit states traveling in
non-inertial frames are investigated, where we assume that all particles
are accelerated. These properties are including fidelities, capacities and
entanglement of the accelerated channels for three different states, namely,
Greeberger-Horne-Zeilinger (GHZ) state, GHZ-like state and W-state. It is shown here that all these properties
are decreased as the accelerations of the moving particles are increased.
The obtained results show that the GHZ-state is the most robust one comparing to the others,
where the degradation rate is less than that for the other states particularly in
the second Rindler region. Also, it is shown here that the entangled property doesn't change in the
accelerated frames. Additionally, the paper shows that the degree of entanglement decreases as
the accelerations of the particles increase in the first Rindler region. However in the second
region, where all subsystems are disconnected at zero acceleration, entangled channels
are generated as the acceleration increases.

\section{Introduction}
Recently, quantum information theory in the relativistic framework
has attracted considerable attention. It seems to be mainly due
to the fact that many modern experiments on quantum information
processing involve the use of photons and/or electrons, where the
relativistic effect is not negligible \cite{Nielsen}.\ The dynamics of
entanglement and its applications for  traveling systems in
non-inertial frames have been investigated by many authors. For
example,  Alsing et al. (2003)  used  the accelerated channel to
perform quantum teleportation  \cite{Milburn}. Landulf et al. (2009)
discussed the phenomena of entanglement sudden death and
information lose for a two qubits system in the non-inertial frames
\cite{Land}. Khan et al. (2011) investigated the relativistic quantum  game in the
non-inertial frames, where the impact of the Unruh effect on the
non-zero sum games was discussed \cite{Khan}.  Metwally (2012) discussed the usefulness of the
travelling channels  in the non-inertial frames to perform effective quantum
teleportation of accelerated and non accelerated information\cite{Metwally,Metwally1}.

Futhermore, the dynamics of tripartite qubits in the non-inertial
frames have been discussed from different point of views. For
example, Hwang et al. (2001) examined the  tripartite
entanglement when one of the three parties moves with a uniform
acceleration with respect to the other parties \cite{Park}. Esfahani et al. (2012)
discussed the dynamics of entanglement for three
spin $\frac{1}{2}$ massive particles by Gaussian momentum
distribution \cite{Nasr}.

In this paper, we investigate the behavior of the tripartite entanglement in the non-inertial frames through  GHZ state, GHZ-like state and W-state,where these states  could be used to perform many tasks of quantum
information. Also, we investigate the fidelity property and the channel capacity property of these three states in the accelerated frames.

The outline of this paper is as follows: An analytical solution for the proposed model for the different
three states is introduced in Sec.2. The behavior of the fidelities and the capacities of
the accelerated channels are described in Sec.3. Quantifying the
degree of entanglement of the accelerated channels is provided in
Sec.4. Finally, Sec.5 concludes this paper.

\section{The Proposed Model}
In this section we investigate the behavior of the three different
types of multi-qubit states, GHZ, GHZ-like and W-states, in non-inertial frames.These tripartite states have been
classified by D\"{u}r et. al \cite{Dur}, such that they can't be
obtained from each other by using local operation and classical
communication. These states have been  widely  as quantum channels
to perform different  quantum information and computations tasks.
For example, Karlesson et al. \cite{Kar} showed that GHZ
state can be utilized to establish teleportation. Due to the new
technology, the GHZ state is used as quantum channel to teleport
quantum correlation be means of what is called entangement
swapping \cite{Bose}.

Assume that we have three users Alice, Bob and Chiral share one of
a pure state in the form:
\begin{eqnarray}\label{eqn}
\ket{\psi_w}&=&\frac{1}{\sqrt{3}}(\ket{0 0
1}+\ket{010}+\ket{100}),
\nonumber\\
\ket{\psi_{G}}&=&\frac{1}{\sqrt{2}}(\ket{000}+\ket{111}),
\nonumber\\
\ket{\psi_{GL}}&=&\frac{1}{2}(\ket{0 01}+\ket{010}+\ket{1
00}+\ket{111}).
\end{eqnarray}
where $\ket{\psi_w},\ket{\psi_{G}}$ and $\ket{\psi_{GL}}$
represent the W, GHZ and GHZ-like states respectively. It is
assumed that these states represent  three qubits state of fermions
particle of mass $m $ moves in the Mikowski space. Let us consider
that these states represent a solution for Dirac equation:
\begin{equation}\label{Dirac}
i\gamma^\kappa(\partial_\kappa-\Gamma_\kappa)\psi+m\psi=0,
\end{equation}
where  $\gamma^\kappa$ represents the Dirac matrices,
$\Gamma_\kappa$ are spinorial affine connections and $\psi$ is a
spinor wave functions \cite{un,Edu}. The transformations from
Minkowski space to Rindler space are given by:
\begin{equation}\label{unr}
\ket{0_k}_M=\cos r\ket{0_k^{+}}_{I}\ket{0_k^{-}}_{II}+
e^{-i\phi}\sin r \ket{1_k^{+}}_{I}\ket{1_k^{-}}_{II}, \quad
\ket{1_k}_M= \ket{1_k^{+}}_{I}\ket{0_k^{-}}_{II},
\end{equation}
where $k=a,b$ and $c$. These transformations divide Rindler
space into two regions $I$ and $II$ for fermions and anti-fermions states
respectively, (for more details see \cite{Als,Metwally}).

It is known that, the density operator for each state in Eq.(\ref{eqn}) is obtained as:
\begin{equation}\label{dens}
{\rho}=\sum_{j}p_j\ket{\psi_j}\bra{\psi_j}
\end{equation}

where $p_j$ is the probability for the qubit to be in state $\ket{\psi_j}$.

\subsection{W-state:}

 The state vector of W-state is given by  ${\psi_w}$ from Eq.(1). Then by using the transformation in Eq.(\ref{unr}), we get the W-state vector in both of Rindler regions $I$ and $II$. By calculating the density operator as in Eq.(\ref{dens}) for the W-state and trace out the anti-fermions particles in the second region $II$, we get the
accelerated channel of the initial W- state, $\rho_w$ in the
first region $I$ as:
\begin{eqnarray}
\rho_{W}^{(I)}&=&\frac{1}{3}\Bigl\{\ket{100}\Bigl(|C_2|^2|C_3|^2\bra{100}+C_2C_1^*|C_3|^2\bra{010}+C_3C_1^*|C_2|^2\bra{001}\Bigr)
\nonumber\\
&&+
\ket{010}\Bigl(C_1C_2^*|C_3|^2\bra{100}+|C_1|^2|C_3|^2\bra{010}+C_3C_2^*|C_1|^2\bra{001}\Bigr)
\nonumber\\
&&+
\ket{001}\Bigl(C_1C_3^*|C_2|^2\bra{100}+C_2C_3^*|C_1|^2\bra{010}+|C_1|^2|C_2|^2\bra{001}\Bigr)
\nonumber\\
&&+
\ket{101}\Bigl(\bigl\{|S_1|^2|C_2|^2+|C_2|^2|S_3|^2\bigr\}\bra{101}+C_2C_1^*|S_3|^2\bra{011}+C_2C_3^*|S_1|^2\bra{110}\Bigr)
\nonumber\\
&&+
\ket{110}\Bigl(\bigl\{|S_2|^2|c_3|^2+|S_1|^2|C_3|^2\bigr\}\bra{110}+C_3C_1^*|S_2|^2\bra{011}+C_3C_2^*|S_1|^2\bra{101}\Bigr)
\nonumber\\
&&+
\ket{011}\Bigl(\bigl\{|C_1|^2|S_3|^2+|C_1|^2|S_1|^2\bigr\}\bra{011}+C_1C_2^*|S_3|^2\bra{101}+C_1C_3^*|S_2|^2\bra{110}\Bigr)
\nonumber\\
&&+\ket{111}\Bigl(|S_2|^2|S_3|^2+|S_1|^2|S_3|^2+|S_1|^2|S_2|^2\Bigr)\bra{111}\Bigr\},
\end{eqnarray}
where $C_i=\cos r_j, S_i=\sin r_j, i=1,2,3$, $j=a,b,c$  and $r_j$
are the accelerations of the accelerated particles. Similarly if
we trace the fermions particles in the  first region, $I$, we
obtain the  state of anti-fermions in the second region $II$ as:
\begin{eqnarray}
\rho_{W}^{(II)}&=&\frac{1}{3}\Bigl\{
\ket{000}\Bigl(|C_2|^2|C_3|^2+|C_1|^2|C_3|^2+|C_1|^2|C_2|^2)\bra{000}\Bigr)
\nonumber\\
&&+\ket{010}\Bigl(\bigl\{|S_2|^2|C_3|^2+|S_2|^2|C_1|^2\bigr\}\bra{010}+S_2S_1^*|C_3|^2\bra{100}+S_2S_3^*|C_1|^2\bra{001}\Bigr)
\nonumber\\
&&+\ket{001}\Bigl(\bigl\{|C_1|^2|S_3|^2+|C_2|^2|S_3|^2\bigr\}\bra{001}+S_3S_1^*|C_2|^2\bra{100}+S_3S_2^*|C_1|^2\bra{010}\Bigl)
\nonumber\\
&&+\ket{101}\Bigl(|S_1|^2|S_3|^2\bra{101}+S_1S_2^*|S_3|^2\bra{011}+S_3S_2^*|S_1|^2\bra{110}\Bigl)
\nonumber\\
&&+\ket{100}\Bigl(\bigl\{|S_1|^2|C_3|^2+|S_1|^2|C_2|^2\bigr\}\bra{100}+S_1S_2^*|C_3|^2\bra{010}+S_1S_3^*|C_2|^2\bra{001}\Bigl)
\nonumber\\
&&+\ket{110}\Bigl(|S_1|^2|S_2|^2\bra{110}+S_1S_3^*|S_2|^2\bra{011}+S_2S_3^*|S_1|^2\bra{101}\Bigl)
\nonumber\\
&&+\ket{011}\Bigl(|S_2|^2|S_3|^2\bra{011}+S_2S_1^*|S_3|^2\bra{101}+S_3S_1^*|S_2|^2\bra{110}\Bigl)
 \Bigr\}
 \end{eqnarray}
\subsection{GHZ-state:}

In the GHZ-state, we assume three users share a three qubits
state of GHZ type. In the Rindler space and tracing out the states
in the second regions, one gets the behavior of $\rho_{G}$ in
the first region as:
\begin{eqnarray}
\rho_{G}^{(I)}&=&\frac{1}{2}\Bigl\{|c_3|^2\bigl\{|c_1|^2(|c_2|^2\ket{000}\bra{000}+|s_2|^2\ket{010}\bra{010})
+|s_1|^2(|c_2|^2\ket{100}\bra{100}+|s_2|^2\ket{110}\bra{110})\bigr\}
\nonumber\\
 &&+|s_3|^2\bigl\{|c_1|^2\bigl(|c_2|^2\ket{001}\bra{001}+|s_2|^2\ket{011}\bra{011}\bigr)
+|s_1|^2\bigl(|c_2|^2\ket{101}\bra{101}+|s_2|^2\ket{111}\bra{111}\bigr)\bigr\}
 \nonumber\\
 &&+c_1c_2c_3\bigl(\ket{000}\bra{111}+\ket{111}\bra{000}\bigl)+\ket{111}\bra{111}
 \Bigl\}.
\end{eqnarray}
To find the density operator $\rho_G$ in the second region
$II$,  we trace out all the modes in the first region $I$.  After
performing this procedure, one gets, the density operator$
\rho_{G}^{(II)}$ which is the same as $\rho_{G}^{(I)}$ except
the last three terms change into
$s_1s_2s_3\bigl(\ket{000}\bra{111}+\ket{111}\bra{000}\bigl)+\ket{000}\bra{000}$.

\subsection{GHZ-like state:}

 The GHZ-like state represents another version of GHZ state, where  it can be constructed either from an
 EPR pair and a single photon or from GHZ state \cite{Yang}.
In the non-inertial frames this state  is transformed into
$\rho_{GL}$ in the first region $I$ as:
\begin{eqnarray}
\rho_{GL}^{(I)}&=&\frac{1}{4}\Bigl\{\ket{001}\{c_1^2c_2^2\bra{001}+c_1c_3c_2^2\bra{100}+
c_2c_3c_1^2\bra{010}+c_1c_2c_3\bra{111}\}
\nonumber\\
&&+\ket{010}\{c_1^2c_3^2\bra{010}+c_3c_2c_1^2\bra{001}+c_3^2c_1c_2\bra{100}+c_1c_3\bra{111}\}
\nonumber\\
&&+\ket{100}\{c_2^2c_3^2\bra{100}+c_3c_1c_2^2\bra{001}+c_3^2c_2c_1\bra{010}+c_2c_3\bra{111}\}
\nonumber\\
&&+\ket{101}\{(s_1^2c_2^2+c_2^2s_3^2)\bra{101}+c_2c_3s_1^2\bra{110}+s_3^2c_2c_1\bra{011}\}
\nonumber\\
&&+\ket{110}\{(s_2^2c_3^2+s_1^2c_3^2)\bra{110}+c_3c_1s_2^2\bra{011}+c_3c_2s_1^2\bra{101}\}
\nonumber\\
&&+\ket{011}\{(c_1^2s_2^2+c_1^2s_3^2)\bra{011}+s_3^2c_1c_2\bra{101}+c_1c_3s_2^2\bra{110}\}
\nonumber\\
&&+\ket{111}\{(s_1^2s_2^2+s_1^2s_3^2+s_2^2s_3^2+1)\bra{111}+c_1c_3\bra{010}+c_1c_2\bra{001}+c_2c_3\bra{100}\}\Bigl\}
\end{eqnarray}
On the other hand, the behavior of $\rho_{GL}$ in the second
region is described by:
\begin{eqnarray}
\rho_{GL}^{(II)}&=&\frac{1}{4}\Bigl\{\ket{000}{(c_1^2c_2^2+c_1^2c_3^2+c_2^2c_3^2+1)\bra{000}+s_1s_3\bra{101}+s_1s_2\bra{110}+s_2s_3\bra{011}}
\nonumber\\
&&+\ket{001}\{(c_2^2s_3^2+c_1^2ss_3^2)\bra{001}+s_1s_3c_2^2\bra{100}+s_3s_2c_1^2\bra{010}\}
\nonumber\\
&&+\ket{010}\{(s_2^2c_3^2+c_1^2s_2^2)\bra{010}+s_2s_3c_1^2\bra{001}+s_2s_1c_3^2\bra{100}\}
\nonumber\\
&&+\ket{100}\{(s_1^2c_2^2+s_1^2c_3^2)\bra{100}+s_1s_2c_3^2\bra{010}+s_1s_3c_2^2\bra{001}\}
\nonumber\\
&&+\ket{101}\{s_1^2s_3^2\bra{101}+s_1s_3\bra{000}+s_1s_2s_3^2\bra{011}+s_3s_2s_1^2\bra{110}\}
\nonumber\\
&&+\ket{110}\{s_1^2s_2^2\bra{110}+s_1s_2\bra{000}+s_1s_3s_2^2\bra{011}+s_2s_3s_1^2\bra{101}\}
\nonumber\\
&&+\ket{011}\{s_2^2s_3^2\bra{011}+s_2s_3\bra{000}+s_3^2s_2s_1\bra{101}+s_1s_3s_2^2\bra{110}\}
 \Bigr\}
\end{eqnarray}

Now, we have obtained the density operators of the accelerated
tripartite channels in the non inertial frames. Next section is devoted to  investigate some properties of these channels as their
fidelities and their capacities in both regions.

\section{Fidelities and Capacities of the Accelerated Channels}
Here, we investigate the behavior of the fidelities and
the capacities of the acceleration channels in both Rindler's
regions.

 It is well known that the fidelity measures
the closeness of the accelerated channel in a specific region to
its initial state. Mathematically, the fidelity
$\mathcal{F}=tr\Bigl\{\rho_{initial}\rho_{final}\Bigr\}$, where
$\rho_{initial}$ is given by the GHZ, GHZ-like state or W-state
and $\rho_{final}$ is given by the corresponding form of these
states in the regions $I$ or $II$. In this context, the
fidelities in the first region are given by:
\begin{eqnarray}
\mathcal{F}_{W}^{(I)}&=&\frac{1}{9}\Bigl\{(c_1^2c_2^2+c_1^2c_3^2+c_2^2c_3^2+2c_1c_2c_3^2+2c_1c_3c_2^2+2c_3c_2c_1^2\Bigr\},
\nonumber\\
\mathcal{F}_{G}^{(I)}&=&\frac{1}{4}\Bigl\{1+c_1^2c_2^2c_3^2+2c_1c_2c_3\Bigr\},
\nonumber\\
\mathcal{F}_{GL}^{(I)}&=&\frac{1}{16}\Bigl\{1+c_1^2(c_2^2+c_3^2)+c_2^2c_3^2+2c_1(c_2+c_3)+2c_2c_3+2c_1c_2c_3^2
\nonumber\\
&&+ 2c_1c_3c_2^2+
2c_2c_3c_1^2+s_1^2(s_2^2+s_3^2)+s_2^2s_3^2\Bigr\}.
\end{eqnarray}

In the first region, while in the second region $II$, these
fidelities are  defined as:

\begin{eqnarray}\label{Fid-II}
\mathcal{F}^{(II)}_{W}&=&\frac{1}{9}\Bigl\{s_1^2(c_2^2+c_3^2)+s_2^2(c_1^2+c_3^2)+s_3^2(c_1^2+c_2^2)+2s_1s_3c_2^2
+2s_2s_3c_1^2+2s_1s_3c_2^2\Bigr\}, \nonumber\\
\mathcal{F}^{(II)}_{\small{G}}&=&\frac{1}{4}\Bigl\{(1+c_1^2c_2^2c_3^2+s_1^2s_2^2s_3^2+2s_1s_2s_3\Bigr\},
\nonumber\\
\mathcal{F}^{(II)}_{\small{GL}}&=&\frac{1}{16}\Bigl\{s_1^2(c_2^2+c_3^2)+s_2^2(c_1^2+c_3^2)+s_3^2(c_1^2+c_2^2)+2s_2s_3c_1^2
+2s_1s_3c_2^2+2s_1s_2c_3^2\Bigl\}. \nonumber\\
\end{eqnarray}

Fig.1 shows the behavior of the fidelities $\mathcal{F}$ for the accelerated
channels in Rindler regions. Fig.1a,
describes the behavior of the three tripartite channels in the
first region $I$. Generally, it is clear that the fidelities
decrease as the accelerations of the moving particles are
increased. The minimum values of these fidelities are approached as
$r\to\infty$. These minimum values depend on the type of the
accelerated channel. It is easy to notice that the lower bound of the fidelity is
bigger for the W-state and smaller for the GHZ and GHZ like state.
However the GHZ state has  a bigger fidelity than that shown
for the GHZ-like state.

\begin{figure}
  \begin{center}
  \includegraphics[width=19pc,height=14pc]{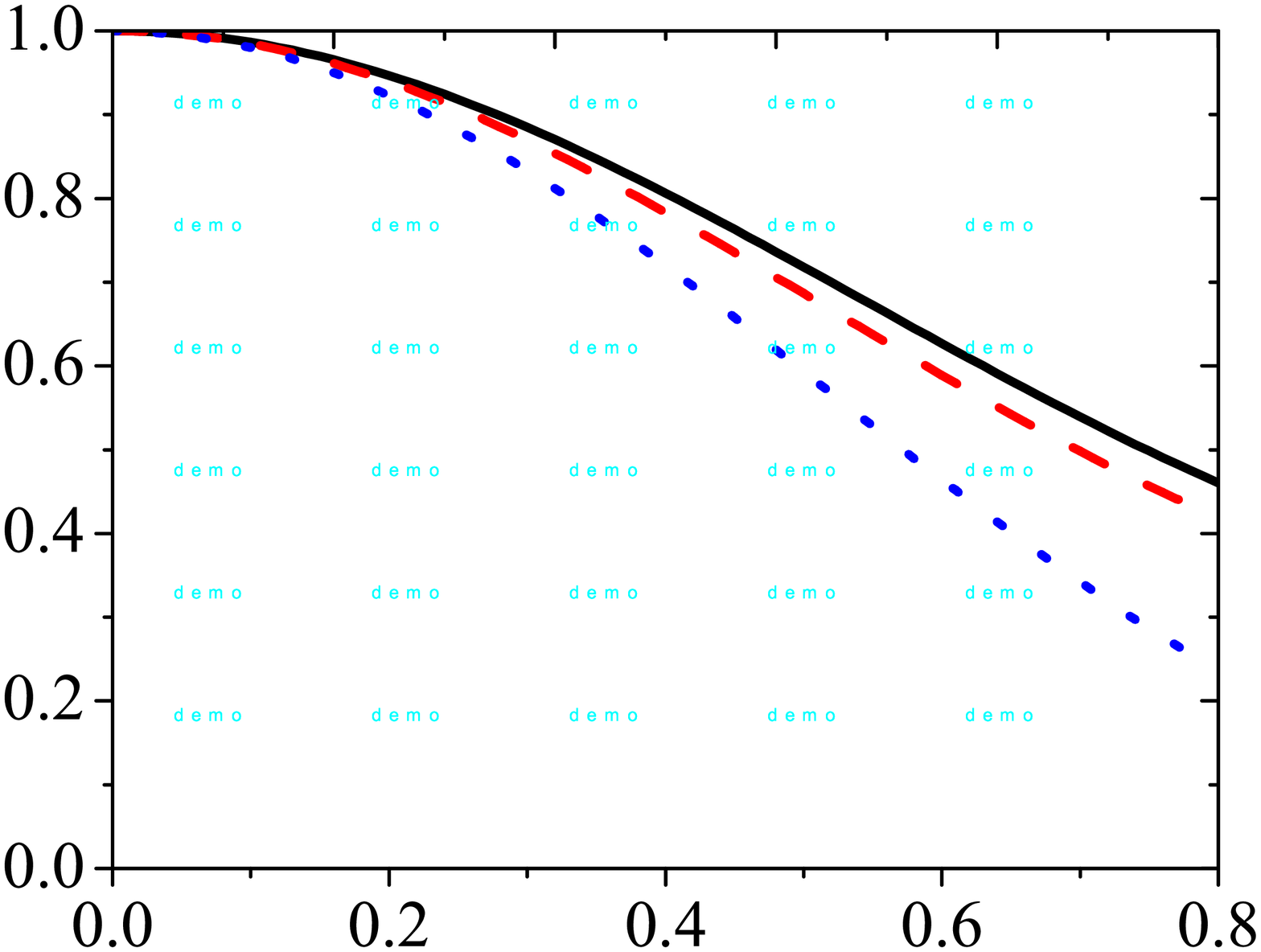}
   \includegraphics[width=19pc,height=14pc]{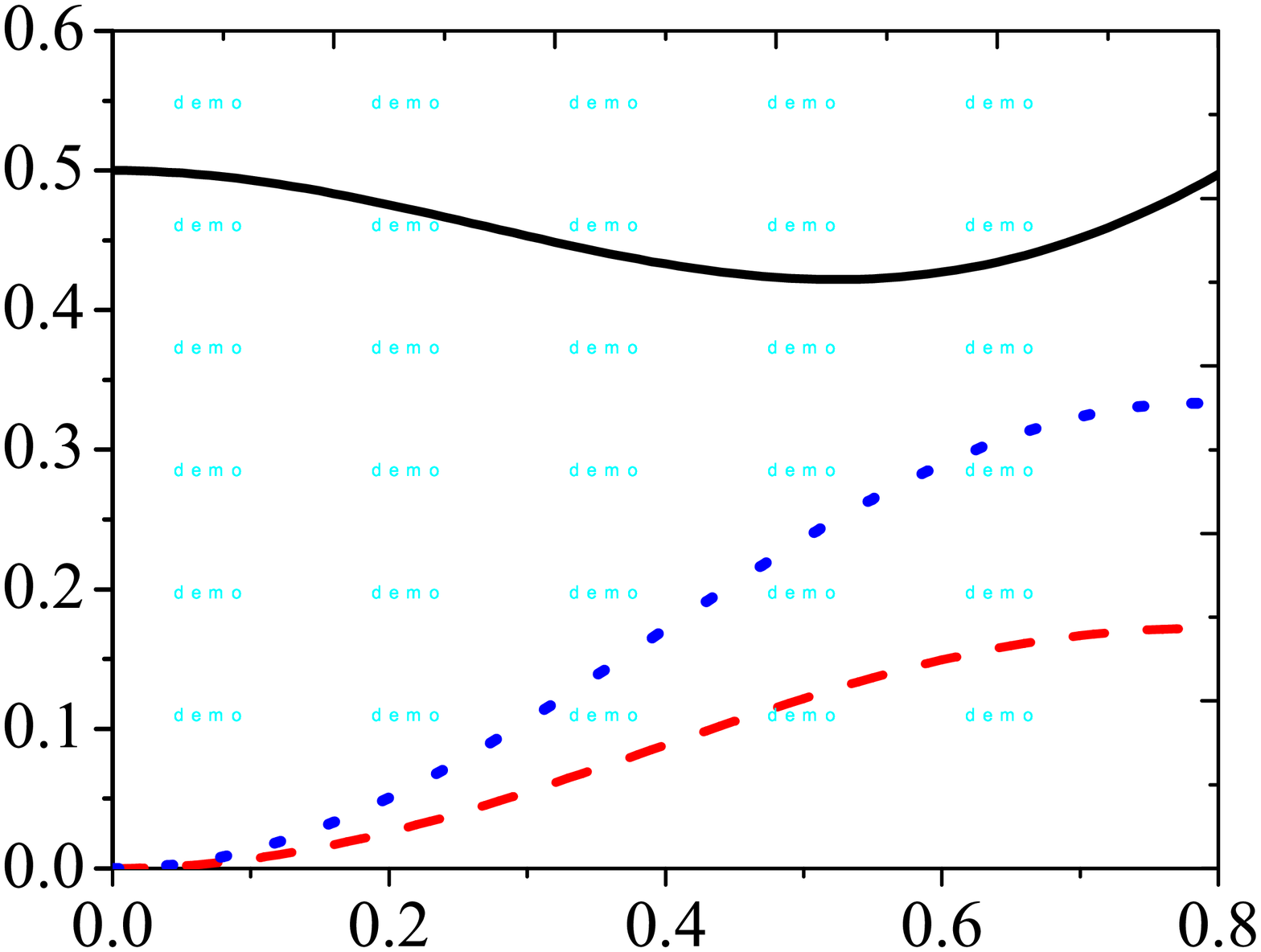}
       \put(-280,135){$(a)$} \put(-50,135){$(b)$}
 \put(-350,5){$r$}
 \put(-100,5){$r$}%
 \put(-470,90){$\mathcal{F}(\rho_i)$}
 \put(-240,90){$\mathcal{F}(\rho_i)$}
       \caption{The fidelities of the accelerated  states in the first   region $I$ .
        The solid, dash and  dot curves represent $\mathcal{F}(\rho_{G}^{(i)})$, $\mathcal{F}(\rho^{(i)}_{GL})$
      and $\mathcal{F}(\rho^{(i)}_{W})$ respectively, $i=I$ and $II$.}
  \end{center}
\end{figure}

The fidelities of the travelling tripartite sates in the second
region $II$ ,which are descibed by (\ref{Fid-II}) are described in Fig.1b.  For the
normal GHZ state the initial fidelity
$\mathcal{F}_{G}^{(II)}=0.5$ at $r_i=0,i=a,b$ and $c$. However, as
the acceleration increases there will be a small change in its
behavior. On the other hand, for the GHZ-like state and the W-state,
the fidelities are zero for zero accelerations. As the
accelerations increase, the fidelities $\mathcal{F}_{GL}^{(II)}$
and $\mathcal{F}_{W}^{(II)}$ increase to reach  their maximum
value as the accelerations tend to $\infty$. The maximum value
 $\mathcal{F}_{GL}^{(II)}$ is bigger than that depicted for
 $\mathcal{F}_{W}^{(II)}$\

\begin{figure}
  \begin{center}
     \includegraphics[width=19pc,height=14pc]{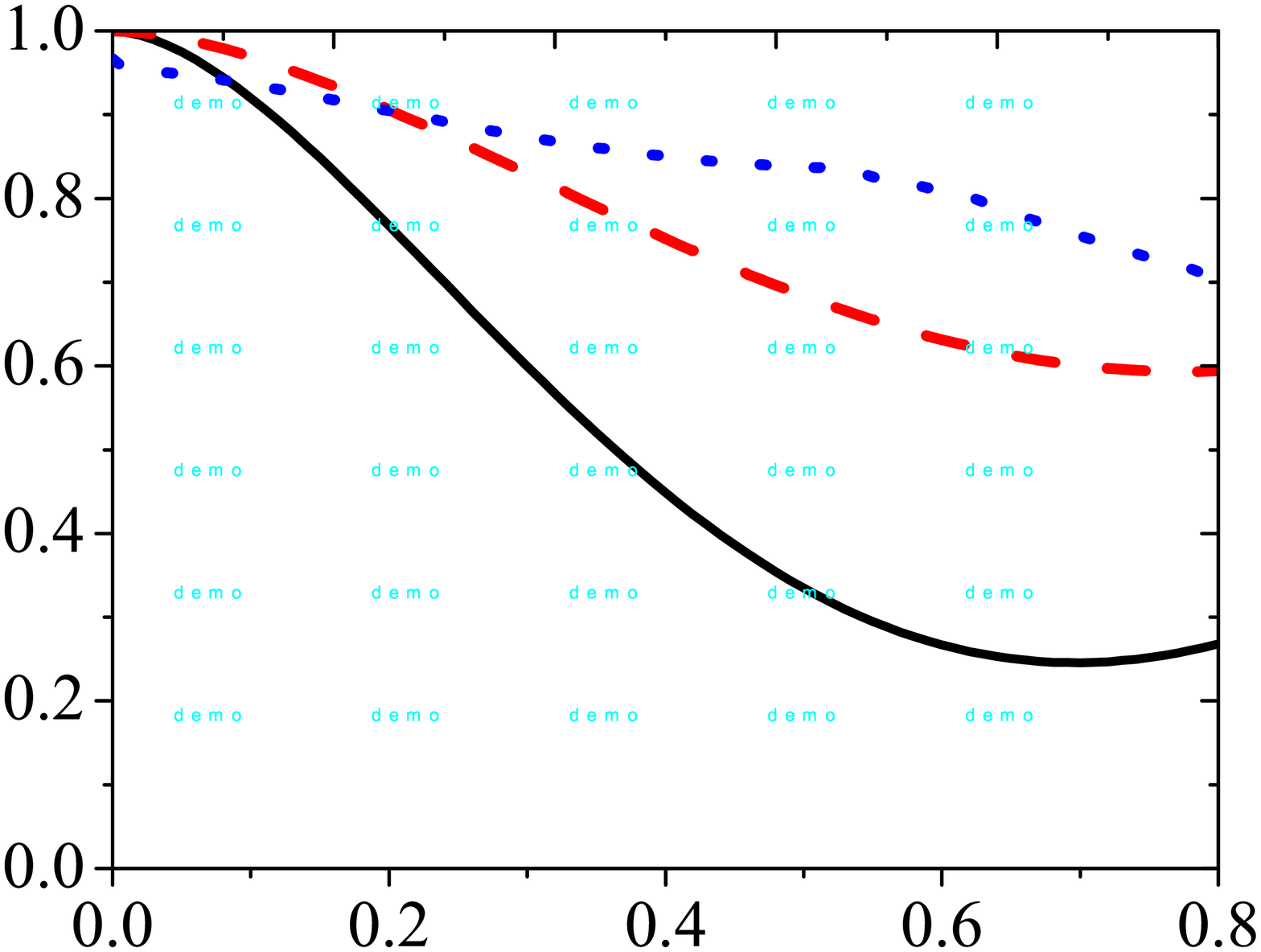}
     \includegraphics[width=19pc,height=14pc]{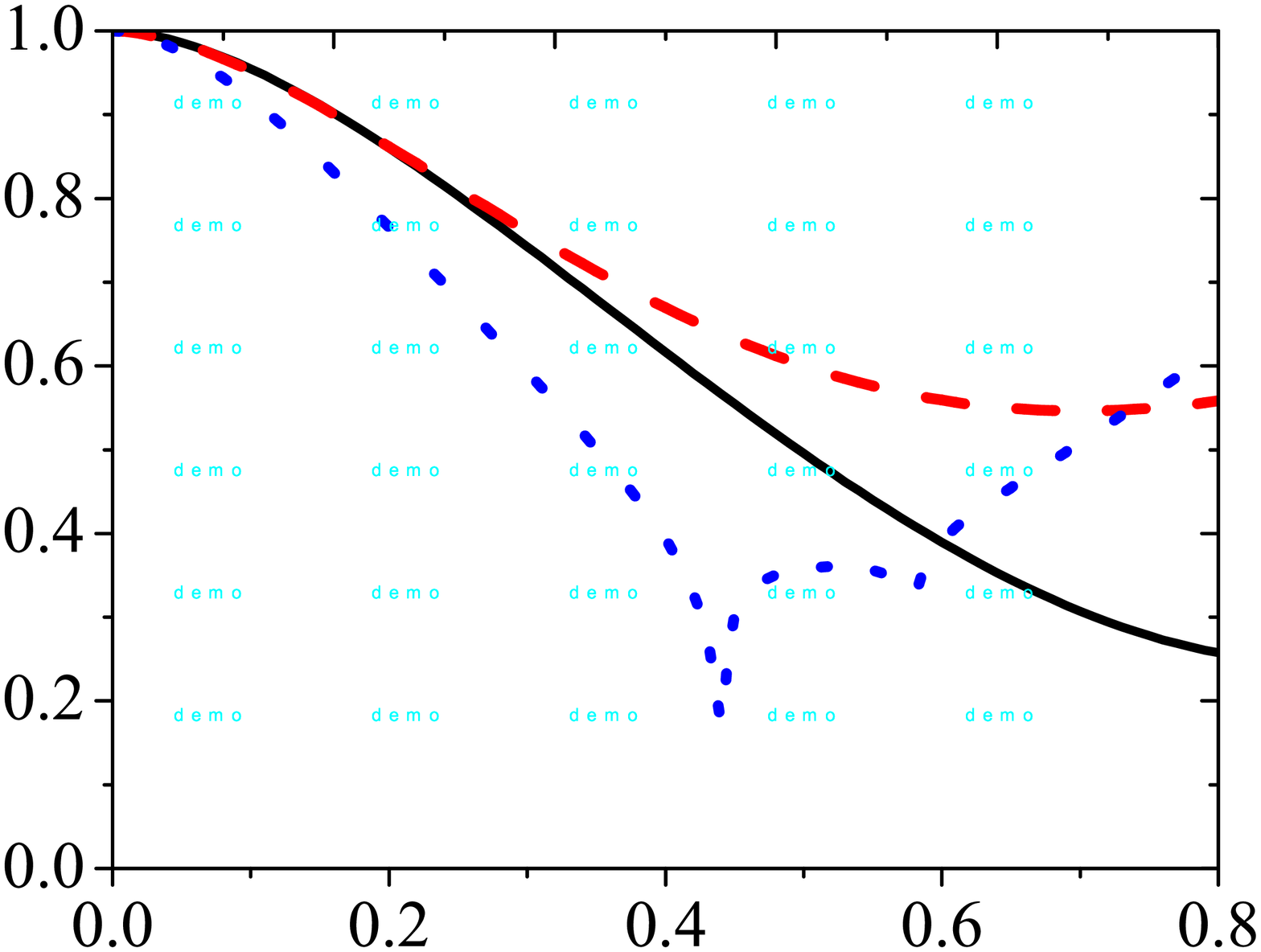}
     \put(-280,135){$(a)$} \put(-50,135){$(b)$}
 \put(-350,5){$r$}
 \put(-100,5){$r$}%
 \put(-460,90){$\bar{\mathcal{C}_p}(\rho_i)$}
 \put(-225,90){$\bar{\mathcal{C}_p}(\rho_i)$}
      \caption{The average capacities $\bar{\mathcal{C}_p}(\rho)$ of the accelerated channels
      in the first Rindler Region (a)$I$ and  the second region (b) $II$. The solid, dash
       and  dot curves represent $\bar{\mathcal{C}_p}({\rho_{G}^{(i)}})$, $\bar{\mathcal{C}_p}(\rho^{(i)}_{GL})$
      and $\bar{\mathcal{C}_p}(\rho^{(i)}_{W})$ respectively, $i=I$ and $II$. }
  \end{center}
\end{figure}
All  quantum information tasks such as quantum computation and
coding information, depend on the capacity of the used quantum
channel. Therefore, it is important to evaluate the transmission
rate of information from a sender to a receiver. For bipartite
state $\rho_{ab}$ the capacity  is given by \cite{Metwally2}:
\begin{equation}
\mathcal{C}_p=log_a D+\mathcal{S}(\rho_b)-\mathcal{S}(\rho_{ab}),
\end{equation}
where  $\rho_b=tr_a\{\rho_{ab}\}$, $D=2$ is the dimension of
$\rho_a$
 and $\mathcal{S}(.)$ is the von Numann entropy. For tripartite
 particles we introduce the average capacity between each
 two particles as a measure of the capacity of the accelerated
 channels. Mathematically, the average  capacity is defined as
 \begin{equation}
\bar{\mathcal{C}_p}(\rho_{abc})=\frac{1}{3}\Bigl(\mathcal{C}(\rho_{ab})+\mathcal{C}(\rho_{ab})+\mathcal{C}(\rho_{ac})\Bigr)
\end{equation}

 Fig.2 shows the behavior of the average  capacities of the  accelerated
channel. Fig.2a describes the dynamics of the
average  capacities in the first Rindler region {\it I},
$\bar{\mathcal{C}_p}(\rho_i^{(I)})$, where $i=G,GL$ or $W$ states.
As a general behavior the capacities decrease as the accelerations
of the particle increase. This means that, the possibility of
coding information decreases as the acceleration increases. The
diminishing rate is bigger for the normal GHZ  state and it is
smaller for the W-state. However in the second region a similar
behavior is depicted in Fig.2b, but the diminishing rate is
smaller than that displayed in Fig.2a.

\section{Entanglement of the Accelerated Channel}
We proceed now for the most important property that we investigate
in this paper. We quantify the degree of entanglement of the
traveling states in the Rindler regions. To pick up the
entanglement features of the bipartite, negativity
$\mathcal{N}(\rho_{abc})$ can be adopted as a measure of the
degree of entanglement  \cite{Vidal,Man,metwally3}. This measure
is defined as
\begin{equation}
\mathcal{N}(\rho_{abc})=2max\bigl\{0,\sum_i|\lambda_i|~\bigr\},
\end{equation}
where $\lambda_i$ are the negative values of the partial transpose
of the density operator $\mathcal{N}(\rho_{abc})$.

The  entanglement  behavior in the first region $I$ of the three
tripartite states is described in Fig.3a, where it is assumed
that the three particles are equally accelerated.  At zero
accelerations i.e. $r_a=r_2=r_3=0$, the fidelity of each state is
maximum in the first region {\it I}, where
$\mathcal{N}_{G}^{(I)}=\mathcal{N}_{GL}^{(I)}=1$, while for the
W-state $\mathcal{N}_{W}^{(I)}\simeq 0.66$. As the acceleration
increases, the entanglement smoothly approachs its minimum values at
$r_i\to \infty, i=1,2,3$. It is clear that the degree of
entanglement of GHZ-like state is smaller than that shown for the
normal GHZ state. However  the minimum value of
$\mathcal{N}_{\rho_{GL}^{(I)}}$ is smaller than that depicted for
$\mathcal{N}_{\rho_{G}}^{(I)}$.
\begin{figure}
  \begin{center}
    \includegraphics[width=19pc,height=15pc]{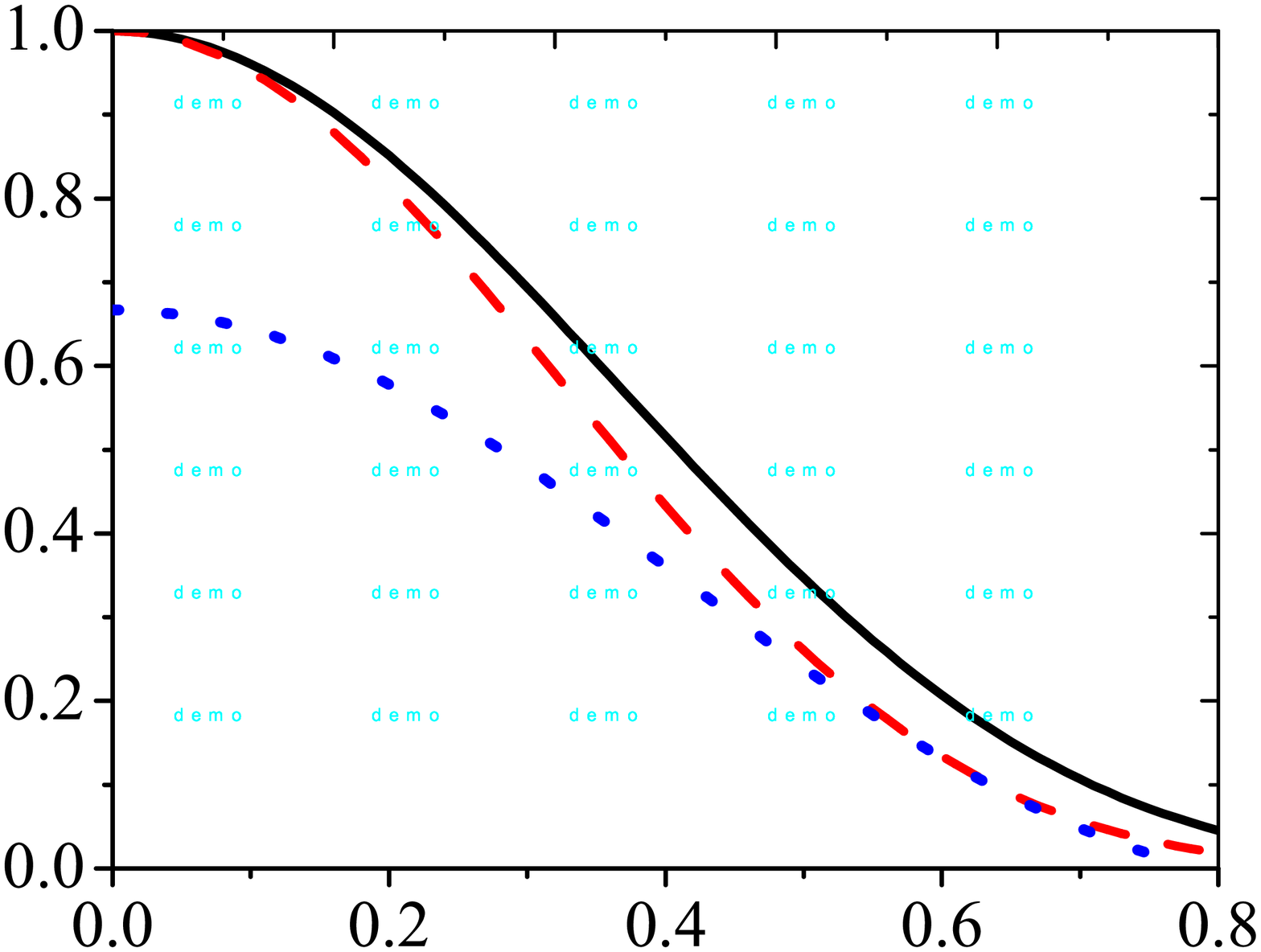}
      \includegraphics[width=19pc,height=15pc]{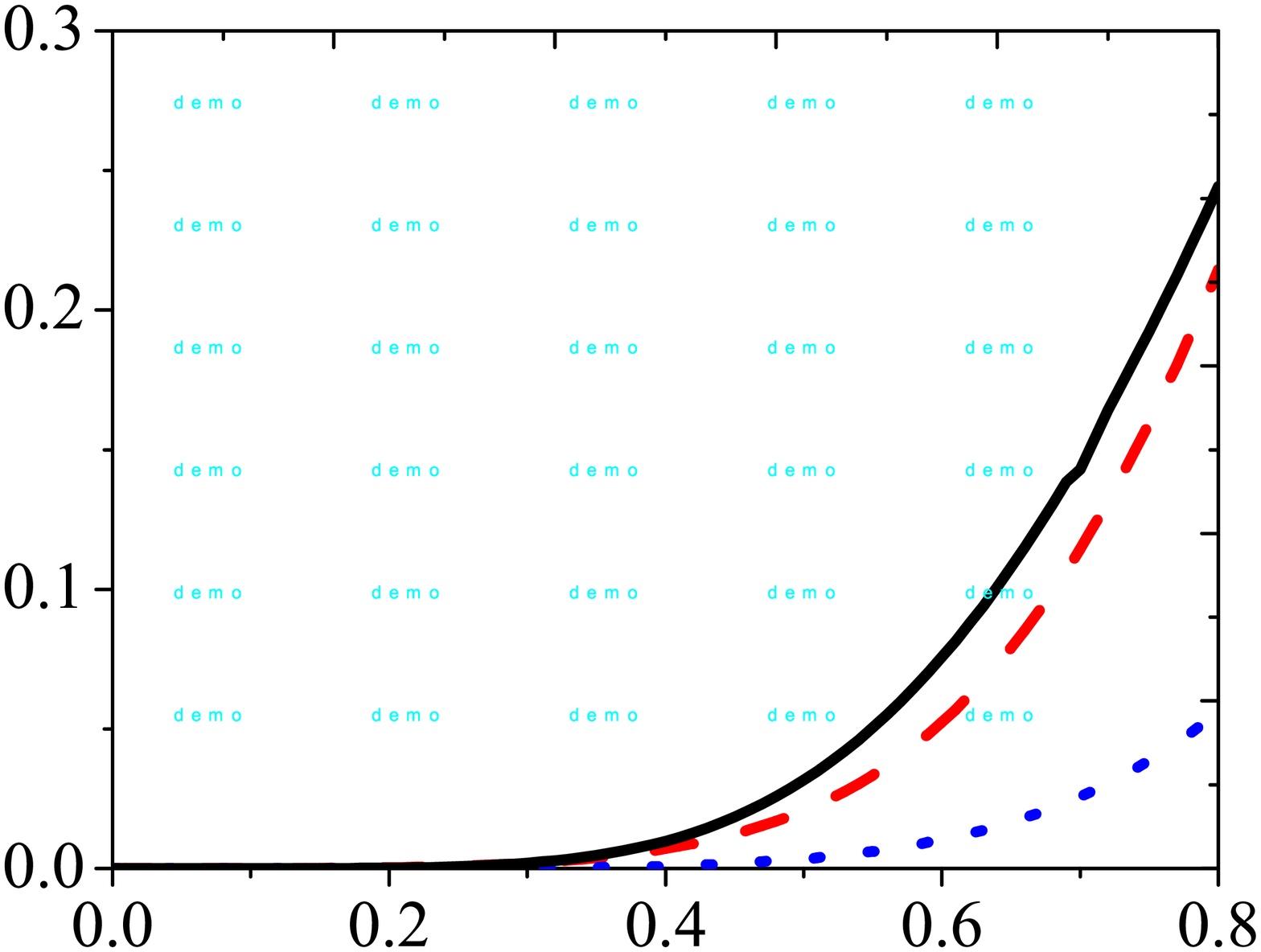}
       \put(-280,140){$(a)$} \put(-50,140){$(b)$}
 \put(-350,5){$r$}
 \put(-100,5){$r$}%
 \put(-460,90){$\mathcal{N}(\rho_i^{(I)})$}
 \put(-235,90){$\mathcal{N}(\rho_i^{(II)})$}
      \caption{The entanglement of  the accelerated channels in the Rindler region(a) $I$ and (b) $II$. The solid, dash
       and  dot curves represent $\mathcal{N}({\rho_{G}^{(i)}})$, $\mathcal{N}(\rho^{(i)}_{GL})$
      and $\mathcal{N}(\rho^{(i)}_{W})$ respectively, $i=I$ and $II$. }
  \end{center}
\end{figure}
The behavior of the W-state is similar  than those shown for GHZ
family and the $\mathcal{N}(\rho_W)$ is completely  vanishes  as
$r_i\to\infty$.

Fig.3b, describes the entanglement of the accelerated channels in
Fig.3b, describes the entanglement of the accelerated channels in
the second region $II$. It is clear that at zero acceleration, all the accelerated channels are separable. As  the accelerations
$r_i$ increase, the accelerated channels turns into entangled
channels. The degree of entanglement approaches its maximum upper
bounds as $r_i\to\infty$

\section{Conclusion}
In this paper, we investigated some properties of GHZ, GHZ-like state and W-state in the non inertial frames, where it is assumed
that all the particles are equally accelerated. Namely, we investigated the fidelities, capacities and entanglement. These
phenomena have been investigated in both Rindler regions.
Analytical expressions have been introduced for these quantities
for the different accelerated channels.

It is shown here that, the fidelities of the channels  in the first
Rindler region decrease as the accelerations of the moving
particles increase. However the behavior of the fidelities in the
second region increase for the GHZ-like state and the W-state, where
the fidelities is completely vanish for small values of the
accelerations. The upper limit of the fidelity of the GHZ-like
state is bigger than that depicted for the W-state. On the other
hand, for the normal GHZ state the fidelity is fluctuated around
$0.5$. This asserts that the normal GHZ state is more robust in the
second region.

The capacities of the accelerated channels are quantified as the
average capacities over all the bipartite subsystems. It is shown
that the capacities decrease as the accelerations increase.
However For the normal GHZ state the degradation is bigger than
those depicted for the GHZ-like state and W-state.  However the
degradation rate in the first Rindler region is bigger than that
shown in the second Rindler region. Also,  for the normal GHZ
state the degradation is the biggest one, while it is the smallest for the
W-state. This confirmd that, it is possible to code information in the case of
W-state better than that for the case of GHZ and GHZ like states.

Quantifying the degree of entanglement of the accelerated channel
in the different Rindler region is quantified by means of
negativity.  In the first rindler's region, the entanglement of
the generated entangled channel decreases as the acceleration of
the moving particle increases. However, the decay rate  of
entanglement for GHZ is  smaller than that depicted  for GHZ-like
state. In the second region, there  are entangled channels are
generated for faster accelerations. The upper limit of the degree
of entanglement is the biggest for the GHZ state, while  for the W-state it
is the smallest.

{\it In conclusion:} the entangled properties of the travelling
channels don't change in the non inertial frames. Although the
W-state has less degree of entanglement, we can coded more information than through the GHZ and the GHZ-like states. The
GHZ state is more robust than GHZ-like state.\\\\
{\bf \large Acknowledgment:}~We would like to thank Prof. Abdelmageed
Aly and Dr. Saleh Aly  for their help and support.

\end{document}